\documentclass[aps,prb,twocolumn,superscriptaddress]{revtex4-2}

\usepackage{float}
\usepackage{amsmath}
\usepackage{mathtools} 
\usepackage{amssymb}
\usepackage{amsthm}
\usepackage{bm} 
\usepackage{upgreek} 
\usepackage{xcolor}
\usepackage{graphicx}
\usepackage{subfigure}
\usepackage{epstopdf}
\usepackage{ulem}
\usepackage[colorlinks,linkcolor=blue,anchorcolor=blue,citecolor=blue,urlcolor=blue]{hyperref}
\usepackage{braket}

\newcommand{\blue}[1]{{\color{blue}#1}}

\newcommand{\nn}{{\nonumber}}

\newcommand{\up}{\uparrow}
\newcommand{\dn}{\downarrow}

\newcommand{\av}[1]{\left\langle #1 \right\rangle}

\newcommand{\me}{\mathrm{e}}
\newcommand{\w}{\omega}

\begin{document}
\title{Charge bond order and s-wave superconductivity in the kagome lattice with electron-phonon coupling and electron-electron interaction}
\author{Qing-Geng Yang} \thanks{Q.G.Y. and M.Y. contributed equally to this work.}
\affiliation{National Laboratory of Solid State Microstructures $\&$ School of Physics, Nanjing University, Nanjing 210093, China}

\author{Meng Yao} \thanks{Q.G.Y. and M.Y. contributed equally to this work.}
\affiliation{National Laboratory of Solid State Microstructures $\&$ School of Physics, Nanjing University, Nanjing 210093, China}

\author{Da Wang} \email{dawang@nju.edu.cn}
\affiliation{National Laboratory of Solid State Microstructures $\&$ School of Physics, Nanjing University, Nanjing 210093, China}
\affiliation{Collaborative Innovation Center of Advanced Microstructures, Nanjing University, Nanjing 210093, China}

\author{Qiang-Hua Wang} \email{qhwang@nju.edu.cn}
\affiliation{National Laboratory of Solid State Microstructures $\&$ School of Physics, Nanjing University, Nanjing 210093, China}
\affiliation{Collaborative Innovation Center of Advanced Microstructures, Nanjing University, Nanjing 210093, China}

\begin{abstract}
The effects of optical bond phonons coupled to electrons in two-dimensional lattices have attracted much interest recently, with the hope to explore unconventional superconducting mechanism and pairing symmetries.
Here we conduct a systematic investigation of such phonon modes in the kagome lattice at and around the upper van Hove filling, in order to unravel new effects of the bond phonons in the presence of the unique sublattice frustration. We
combine the singular-mode functional renormalization group and the projector determinant quantum Monte Carlo methods.
At the upper van Hove filling and in the absence of the Hubbard interaction $U$, we find there exists an s-wave superconducting state at weaker electron-phonon coupling constant $\lambda$ and higher phonon frequency $\w$, and a charge bond order (or the valence bond solid) state at larger $\lambda$ and lower $\w$.
The Hubbard interaction $U$ suppresses drastically the s-wave pairing, so that only the charge bond order survives.
On the other hand, upon slight doping away from the van Hove filling, we observe that the charge bond order is suppressed due to the breakdown of the perfect Fermi surface nesting, while the superconductivity persists.
The s-wave superconductivity and charge bond order may be relevant in the layered kagome superconductors AV$_3$Sb$_5$ (A=K, Rb, Cs).
\end{abstract}
\maketitle

\section{INTRODUCTION}
The two-dimensional optical Su-Schrieffer-Heeger (SSH) electron-phonon interaction has attracted much attention in recent years \cite{feldbacher2003coexistence, xing2021quantum, cai2021antiferromagnetism, cai2022, yqg2022, gotze, feng2022, fer_prb_2022, yh2023, zhangprx, han2023, Yirga, XBprb, tanprb, cai2023hightemperature, mar2023}.
On the square lattice, this electron-phonon coupling was found to induce the typical antiferromagnetic (AFM) order \cite{cai2021antiferromagnetism}, whose fluctuations are believed to be strongly related to the unconventional pairing mechanism in high-temperature superconductors like cuprates and iron pnictides \cite{rmp_SC_2012}.
At half filling and in the absence of the Hubbard interaction, the AFM state in the SSH model on a bipartite lattice is actually exactly degenerate with the charge-density-wave (CDW) and s-wave superconductivity (sSC) due to an enlarged SU(4) symmetry in the presence of particle-hole symmetry \cite{feldbacher2003coexistence, yqg2022}.
Upon doping away from half-filling, the spin and charge density waves are suppressed in favor of either sSC for small $U$, or d-wave superconductivity for large $U$ \cite{yqg2022,zhangprx,XBprb,tanprb, yh2023,cai2023hightemperature}.
These studies suggest new avenues for unconventional superconductivity.
The bipartite lattice studied so far enjoys (at half filling) (i) the particle-hole symmetry, and (ii) the van Hove singularity and perfect nesting. While the second type of features are rather common in various types of lattices, the particle-hole symmetry relies on the lattice structure. We ask how the physics changes if the particle-hole symmetry is absent.

\begin{figure}
	\includegraphics[width=\linewidth]{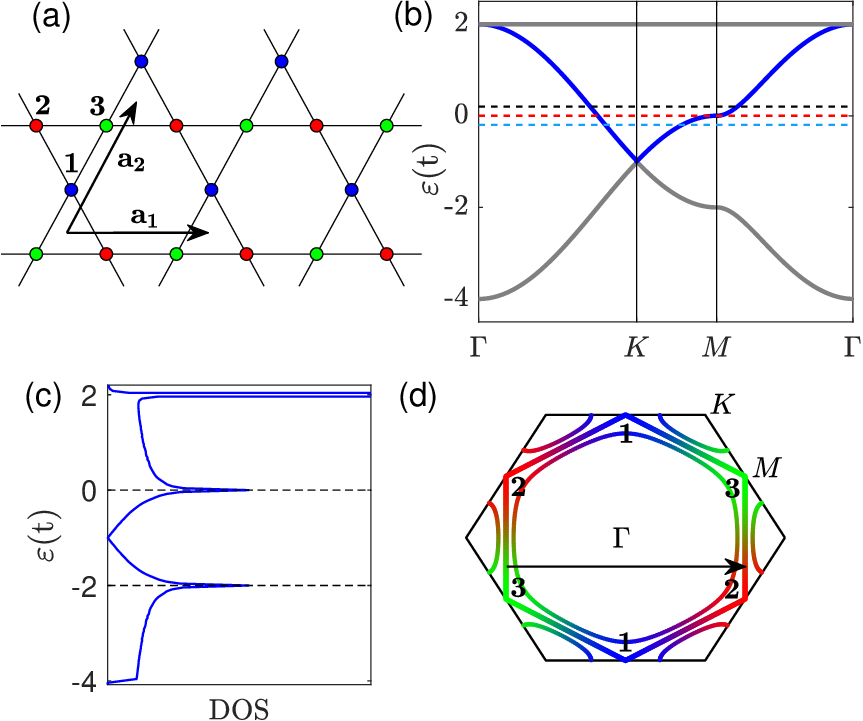}
	\caption{(a) Lattice structure of the kagome lattice, where $\mathbf{a}_1=(1,0)$, $\mathbf{a}_2=(1/2,\sqrt{3}/2)$ denote two primitive translation vectors, and indices $1$, $2$ and $3$ denote different sublattices. (b) Band dispersion along a high symmetry path $\Gamma-K-M-\Gamma$. The black, red and blue dashed lines are fermi levels for $\mu=+0.2$, $0$, $-0.2t$, respectively. (c) Density of states with the two van Hove filling levels indicated by two dashed lines. (d) Fermi surfaces for $\mu=0,\pm0.2t$ with color-scaled sublattice components. The arrow indicates the nesting vector at the van Hove filling.}
	\label{fig:band}
\end{figure}

Here we focus on the two-dimensional kagome lattice which does not have particle-hole symmetry in the band structure. It is particularly interesting because of its many intriguing properties \cite{Yin_N_2022}.
The inherent lattice frustration leads to the emergence of spin disordered phases, such as quantum spin liquid or valence bond solid (VBS), as obtained from extensive studies on the underlying quantum spin model \cite{Zeng,Sachdev_PRB_1992,Syr,Wang_PRB_2006,WXG,singh,Yan_S_2011,Wan_PRB_2013,Zhou_RMP_2017} in the Mott limit.
On the other hand, in the itinerant limit, its electronic band structure exhibits three appealing ingredients simultaneously: Dirac cones, flat band, and van Hove singularities, as shown in Fig.~\ref{fig:band}.
The Dirac cones at the $K$- and $K'$-points have attracted much attention in view of various topological properties such as strongly correlated Dirac metal \cite{Mazin}, non-trivial Chern bands \cite{Yin_2020,Do}, and intrinsic quantum anomalous Hall effect \cite{Xu,Li-2021}.
The flat band at the band top has been customarily taken to study the intriguing emergent phenomena like magnetism \cite{flat-6,flat-3,flat-5,flat-1,flat-7} and high temperature fractional quantum Hall effect \cite{flat-4,flat-2}.
Besides, by tuning the electron filling level to $2/3\pm1/6$ per site, the Fermi surface touches the van Hove points $M$ and exhibits perfect nesting.
At the upper van Hove filling level $5/6$, the Fermi surface is plotted in Fig.~\ref{fig:band}(d) as the middle hexagon-shaped contour. The colors encode the sublattice components, which varies significantly on the contour. This can lead to the matrix element effect in the quasiparticle scattering.\cite{matrix-1,matrix-2,wang2013competing,matrix-3,matrix-4}.
The combination of the van Hove singularity, the perfect nesting and the matrix element effect could lead to various electronic instabilities under the Coulomb interactions \cite{matrix-1,matrix-2,wang2013competing,matrix-3,matrix-4}.

\begin{figure}
	\includegraphics[width=\linewidth]{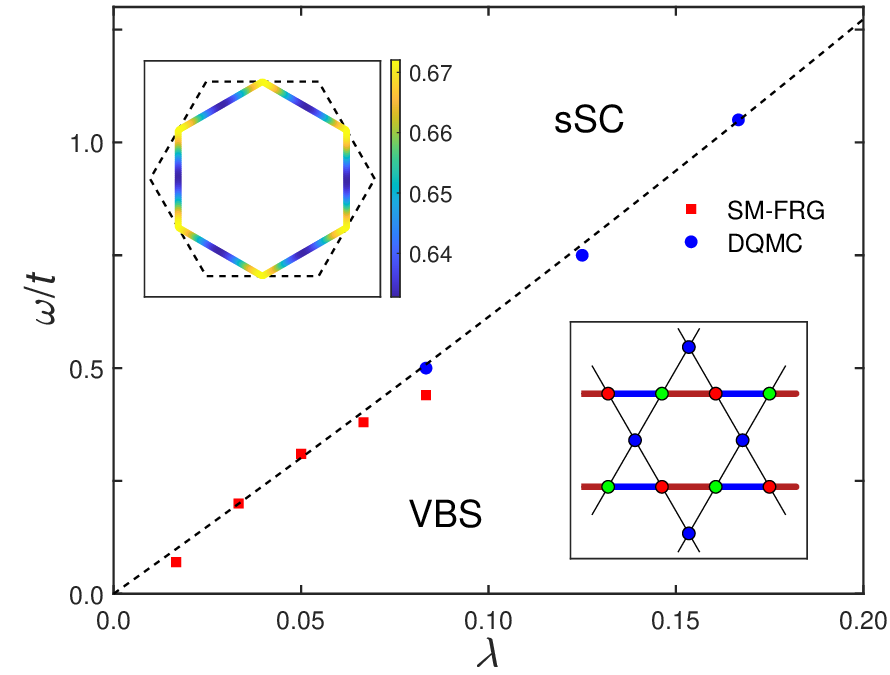}
	\caption{The phase diagram of the SSH model on the kagome lattice at upper van Hove filling obtained from SM-FRG and DQMC. The dashed line is a schematic plot of the phase boundary between sSC and VBS. The upper left inset shows the SC gap function on the Fermi surface. The lower right inset shows the configuration of the VBS state with red/blue colors representing strong/weak bonds.}
	\label{fig:phase}
\end{figure}

In comparison to square lattices with particle-hole symmetry, here we conduct a systematic investigation of the effects of SSH phonon in the kagome lattice at and around the upper van Hove filling by combining the singular-mode functional renormalization group (SM-FRG) \cite{wang2013competing, wang2015phonon,yqg2022} and the projector determinant quantum Monte Carlo (DQMC) \cite{dqmc_method} methods. At the upper van Hove filling, we find sSC exists for smaller electron-phonon coupling constant $\lambda$ and higher phonon frequency $\w$, while increasing $\lambda$ or lowering $\w$ drives the system into a charge bond order (or the VBS) state, with the ordering momenta corresponding to the nesting vectors. These results are summarized as a phase diagram shown in Fig.~\ref{fig:phase}.
Moreover, we further study the effect of the repulsive Hubbard-$U$ and the effect of doping away from the van Hove filling using SM-FRG. We discover that a small Hubbard $U$ drastically suppresses the SC and drives the system into the VBS state at half filling, and in the doped system only the sSC state survives.

\section{MODEL} \label{sec:model}

We consider the optical SSH phonon on each nearest-neighbor bond in the kagome lattice. The total Hamiltonian reads $H=H_0+H_{\rm ph}+H_{\rm eph}$ with the three terms describing electrons ($H_0$), phonons ($H_{\rm ph}$), and electron-phonon coupling ($H_{\rm eph}$), respectively. These three terms are explicitly given by
\begin{align}\label{hamiltonian}
H_0 &= -t\sum_{\av{ij}\sigma}(c^\dagger_{i\sigma}c_{j\sigma}+\mathrm{H.c.}) - \mu\sum_in_i ,\\
H_{\rm ph} &= \omega\sum_{\av{ij}}\left( b_{ij}^\dag b_{ij}+\frac12\right) , \\
H_{\rm eph} &= \frac{g}{\sqrt{2M\omega}} \sum_{\av{ij}\sigma}(b_{ij}+b_{ij}^\dag)(c^\dagger_{i\sigma}c_{j\sigma}+{\rm H.c.}) ,
\end{align}
where $t$ is the hopping integral on each nearest neighbor (NN) bond $\langle ij \rangle$, $c^\dagger_{i\sigma}$ creates an electron at site $i$ with spin $\sigma$ ($\up$ or $\dn$), $\mu$ is the chemical potential, and $n_i=\sum_{\sigma} c_{i\sigma}^\dagger c_{i\sigma}$ is the local electron density.
$b^\dag_{ij}$ creates an optical phonon with single-frequency $\w$ on each NN bond $\langle ij \rangle$, which couples to the electrons in the charge bond channel with the coupling strength $g$, and $M$ is the mass of the vibrating atom.
Note that $M\omega^2=K$ is just the spring constant for the optical phonon mode. In the following, we take $t$ as the energy unit and define the dimensionless electron-phonon coupling constant $\lambda=g^2/(KW)$ as usual.

After integrating out the phonon degrees of freedom, we obtain a retarded electron-electron attraction $V_{\rm eff}=\frac12B_{ij} \Pi B_{ij}$ where $B_{ij}=\sum_\sigma (c_{i\sigma}^\dag c_{j\sigma}+H.c.)$ and $\Pi = -\lambda W \omega^2/(\omega^2+\nu^2)$ for each NN bond, with
$W=6t$ is the bare bandwidth and $\nu$ is the bosonic Matsubara frequency.
(Strictly speaking, the fields in $B_{ij}$ should be understood as Grassmann fields in the path-integral representation of the system.)
Clearly, at $\nu=0$, $V_{\rm eff}$ becomes a direct attraction in the charge bond channel, hence, favors the VBS order. On the other hand, $V_{\rm eff}$ has an attractive component in the local sSC pairing channel as well. These two effects compete with each other and lead to the phase diagram shown in Fig.~\ref{fig:phase}, as we will present in details below.

\section{RESULTS AND DISCUSSIONS} \label{sec:results}

\begin{figure}
	\includegraphics[width=\linewidth]{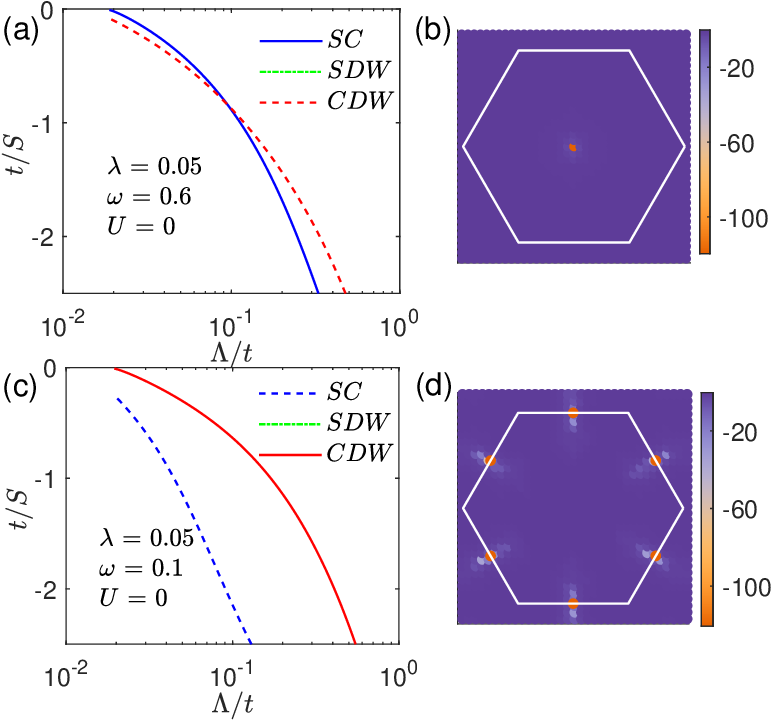}
	\caption{(a) FRG flows of the negative leading singular values $S$ in the SC and CDW channels at $\lambda=0.05$ and $\w=0.6$. The result of the SDW channel is too weak to be seen. (b) plots $S(\mathbf{q})$ in the divergent SC channel, with the white line denoting the Brillouin zone.
    (c) shows the FRG flows at $\lambda=0.05$ and $\w=0.1$, with the corresponding $S(\mathbf{q})$ in the divergent CDW channel plotted in (d). }
\label{fig:sflow}
\end{figure}

We combine SM-FRG \cite{wang2013competing, wang2015phonon,yqg2022} and projector DQMC \cite{dqmc_method} to study the SSH model on the kagome lattice. In principle, SM-FRG is reliable in weak and moderate coupling regimes, while the projector DQMC is more suitable for moderate and strong couplings. Therefore, combining these two methods enables us to draw a convincing conclusion.

\subsection{SM-FRG results} \label{subsec:smfrg}

In SM-FRG, the one-particle-irreducible (1PI) four-point interaction vertex is computed versus a running (lowering) infrared cutoff energy scale $\Lambda$ (the smallest fermionic Matsubara frequency in our case).
From the same 1PI vertex at a given $\Lambda$, we can extract the effective interactions
$V_{\rm{SC}}$, $V_{\rm{SDW}}$ and $V_{\rm{CDW}}$, respectively. They may be understood as scattering matrices between fermion bilinears in the respective channels. In this way, the three channels are treated on equal footing, which is crucial in strongly correlated systems where various ordering tendencies compete closely.
We trace the negative leading eigenvalues $S$ of these scattering matrices during the FRG flow until the largest one diverges at a critical energy scale $\Lambda_c$. This divergence indicates an instability towards the formation of a long-range order in the form of the associated eigen scattering mode, {and $\Lambda_c$ serves as a measure of the transition temperature $T_{c}\sim\Lambda_{c}$}.
More technical details of the SM-FRG algorithm can be found in the appendix and references \cite{wang2013competing, wang2015phonon,yqg2022}.

Fig.~\ref{fig:sflow}\blue{(a)} shows a typical FRG flow of the negative leading singular values $S$ in the SC and CDW channels at $\lambda=0.05$ and $\w=0.6$. Note that $t/S$ is plotted such that $S$ diverges as $t/S$ grows up to zero. The value in the SDW channel is too weak to be seen in the view window.
At high energy scales, the CDW channel dominates, indicating strong charge fluctuations.
With decreasing $\Lambda$, both CDW and SC channels grow up quickly, but the SC channel grows up faster, bypasses the CDW channel and finally diverges at first, indicating the instability in the SC channel.
In Fig.~\ref{fig:sflow}(b), we plot the negative leading eigenvalue $S(\0q)$ as a function of the Cooper pair momentum at the critical energy scale $\Lambda_c$. Clearly, $S(\0q)$ is maximal at zero momentum, corresponding to the Cooper pairing.
By examining the leading eigen scattering mode in the SC channel at $\Lambda_c$, we find the SC is dominated by on-site pairing, hence, corresponding to s-wave pairing.
To see that explicitly, we plot the gap function on the Fermi surface in the left inset of Fig.~\ref{fig:phase}, where the s-wave symmetry is clear. The mild variation in amplitude is because of sublattice mixing.
{The s-wave SC symmetry is similar to the one obtained by Holstein phonon \cite{wu2022crossover} and first-principle calculations \cite{DFT_PRB_2023}. But our FRG result for the SSH phonon reveals its strong competition with charge orders as shown below.
}

On the other hand, there is another possibility of the FRG flow as plotted in Fig.~\ref{fig:sflow}\blue{(c)} at $\lambda=0.05$ and $\w=0.1$, for which the CDW channel always dominates and diverges at first, indicating the instability in the CDW channel.
The negative leading eigenvalue $S(\0q)$ as a function of the CDW momentum is plotted in Fig.~\ref{fig:sflow}(d). The peaks are at $M$ points, which are exactly the nesting vectors $\bm{Q}=(0,2\pi/\sqrt{3})$, up to reciprocal lattice vectors shown in Fig.~\ref{fig:band}(d).
By further checking the leading eigen scattering mode in the CDW channel, we find the CDW is featured by fermion bilinears on the NN bonds in the form of the VBS state.
The bond strength depends on the ordering momentum, as explicitly shown in the right inset of Fig.~\ref{fig:phase}, where the red and blue colors denote the strong and weak bonds, respectively. In the ordered state, the three degenerate patterns can combine into a star-of-David pattern.
{Here, for the charge order, the onsite-component of the fermion bilinears is found to be much smaller than the NN bond-component. This is consistent with the matrix element effect that the nesting vectors mainly connect different sublattices as depicted in Fig.~\ref{fig:band}(d), hence, disfavoring the onsite charge order.}

For a series of $\lambda<0.1$, we vary $\w$ to determine the phase boundary $\w_c$ between the VBS and sSC. The results are shown (red dots) in the phase diagram Fig.~\ref{fig:phase}.
The phase boundary is roughly linear, $\w_c\propto\lambda$.
The VBS lives in the larger-$\lambda$ and smaller-$\w$ side, while the sSC lives in the other side. This can be roughly understood as follows.
Since $\w$ provides an electron energy cutoff to feel the attractive pairing interaction as in the BCS theory, a higher $\w$ is expected to favor SC, known as the isotope effect.
On the other hand, the retarded electron-electron interaction always contributes an attractive component directly for the VBS without any constraint to the fermion energy scale, hence, would always favor the VBS unless SC emerges first.

The same calculation could be performed for stranger coupling. However, the divergence appears too soon, and this invalidates the FRG based on the truncation at the level of four-point vertices. For this reason, in the following, we resort to DQMC for larger $\lambda$.

\subsection{DQMC results} \label{subsec:dqmc}

\begin{figure}
	\centering
	\includegraphics[width=\linewidth]{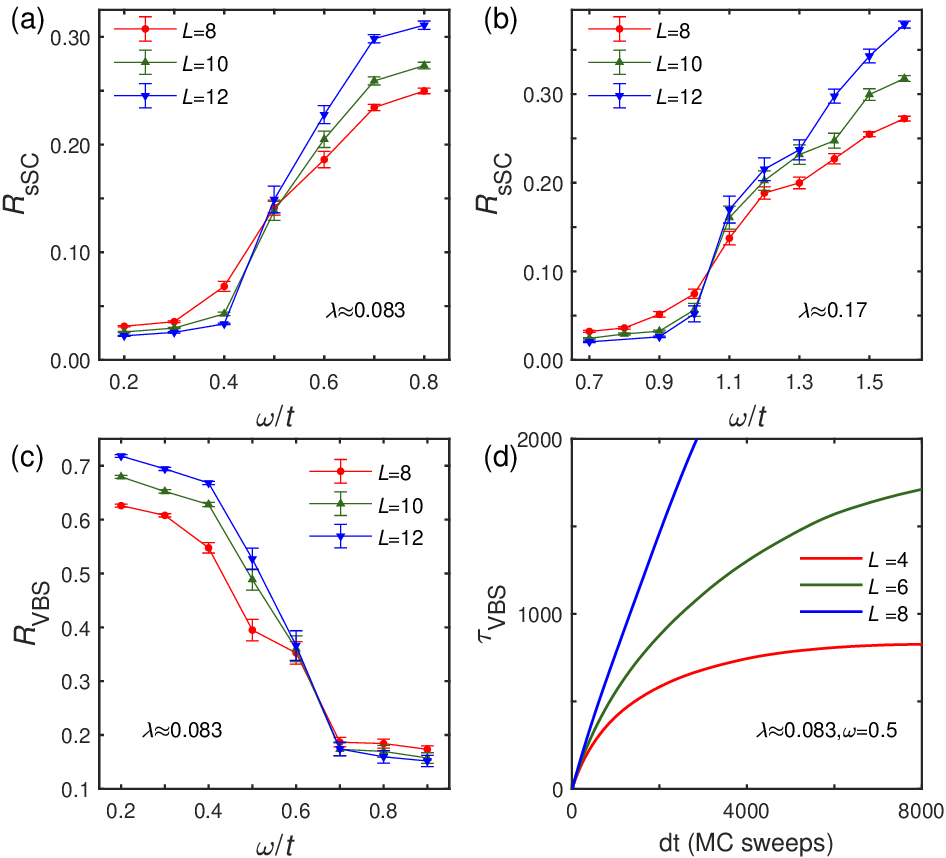}
	\caption{The dimensionless correlation ratios for sSC obtained by DQMC at $L=8,10,12$ are plotted with respect to $\w$ for $\lambda=1/12$ ($\approx0.083$) in (a) and $\lambda=1/6$ ($\approx0.17$) in (b), respectively. The results for VBS at $\lambda=1/12$ are plotted in (c). The poor data quality is attributed to the long autocorrelation time as depicted in (d) (see further discussions in main text).}
	\label{fig:dqmc_1}
\end{figure}

For the SSH bond phonon model, there is no sign problem since spin up and down electrons contribute the same real determinant in the projector DQMC.
By acting the projection $\me^{-\Theta H}$ on a trial state, we obtain the ground state.
To investigate various possible electronic instabilities, we calculate the structure factor (equal-time correlation) $S_{\hat{O}}(\bm{Q},L)=(1/L^4)\sum_{\bm{r},\bm{r'}}\text{e}^{\text{i}\bm{Q}\cdot(\bm{r-r'})} \left( \langle \hat{O}_{\bm{r}}\hat{O}_{\bm{r'}}^{\dagger}\rangle - \langle \hat{O}_{\bm{r}}\rangle\langle\hat{O}_{\bm{r'}}^{\dagger}\rangle\right)$ for different lattice sizes with $L\times L$ unit cells, where $\bm{r}$, $\bm{r}'$ run over the whole lattice, and $\hat{O}$ is a fermion bilinear operator serving as the candidate of a long-range order with ordering momentum $\bm{Q}$.
After obtaining $S_{\hat{O}}(\bm{Q},L)$, we calculate the dimensionless correlation ratio
$R_{\hat{O}}(\bm{Q},L)=1-[S_{\hat{O}}(\bm{Q}+\delta \bm{q},L)/S_{\hat{O}}(\bm{Q},L)]$, where $\delta\bm{q}=\mathbf{b}_{1,2}/L$ is the minimal momentum discretization ($\0b_{1,2}$ are two reciprocal lattice vectors).
As usual, $R_{\hat{O}}\rightarrow 1$ for $L\rightarrow \infty$ implies the formation of a long-range order, while $R_{\hat{O}}\to 0$ gives the opposite case.
Motivated by the above SM-FRG results, we studied both sSC and VBS.
For sSC, $\bm{Q}=0$ and $\hat{O}_{\bm{r}}=c_{\bm{r}\uparrow}^{\alpha}c_{\bm{r}\downarrow}^{\alpha}$ with $\alpha=1,2,3$.
For VBS, $\bm{Q}=(0,2\pi/\sqrt{3})$ and $\hat{O}_{\bm{r}}=\frac12\sum_\sigma \left(c_{\bm{r}\sigma}^{2\dagger}c_{\bm{r}\sigma}^{3}-c_{\bm{r}\sigma}^{3\dagger}c_{\bm{r}+\mathbf{a}_1\sigma}^{2}+H.c.\right)$.
In practice, after carefully checking, we choose the projection time $\Theta$ proportional to $L$ \cite{berg2012}, namely $\Theta=8\,(16)\times L$ for $\lambda=1/6\,(1/12)$, and the Trotter decomposition time slice $\Delta\tau=0.1$ ($0.2$) for $\lambda=1/6$ ($1/12$). 

For sSC, our results of the dimensionless correlation ratio $R_{\rm sSC}$ versus $\w$ are plotted in Fig.~\ref{fig:dqmc_1}(a) for $\lambda=1/12$ and in Fig.~\ref{fig:dqmc_1}(b) for $\lambda=1/6$.
Both plots exhibit data crossings for different lattice sizes $L=8,10,12$, indicating the sSC long-range order is developed for $\w>\w_c$, with $\w_c\approx0.5,1.05$ for the two values of $\lambda$, respectively. In addition, we have also performed simulations at $\lambda=1/8$, giving $\w_c\approx 0.76$ (not shown).
These DQMC data are added to the phase diagram Fig.~\ref{fig:phase}. We find they are in good agreement with the SM-FRG results and broadly extend the linear phase boundary $\w_c\propto\lambda$ to the larger $\lambda$ regime.

For VBS, in Fig.~\ref{fig:dqmc_1}(c) we plot the correlation ratio $R_{\rm VBS}$ versus $\w$ for $L=8,10,12$. The data crossing can be roughly seen, but seems to be a little higher than that of the sSC.
This may imply the coexistence between sSC and VBS near the phase boundary. However, this may also be caused by the severe ergodicity problem even though the global update has been efficiently implemented \cite{Scalettar_PRB_1991}. To see this more clearly, in Fig.~\ref{fig:dqmc_1}(d), we plot the integrated autocorrelation time \cite{autocorrelation} of $S_{\rm VBS}(\bm{Q})$, the saturation value versus the DQMC sweeps gives a practical estimation of the autocorrelation time $\tau_{\rm VBS}$.
Clearly, $\tau_{\rm VBS}$ is already $\sim2000$ DQMC sweeps for $L=6$ and grows up very rapidly (exponentially) with further increasing $L$.
With such long autocorrelation time (which is even much longer for larger $\lambda$, say $1/6$), it is hard for us to obtain high quality data, which may explain the poor data crossing presented in Fig.~\ref{fig:dqmc_1}(c).
Instead, the autocorrelation time for sSC is found to be quite small (tens to hundreds of DQMC sweeps, not shown) such that the above sSC results are still reliable.

\subsection{Effect of Hubbard $U$} \label{hubbard}

\begin{figure}
	\includegraphics[width=\linewidth]{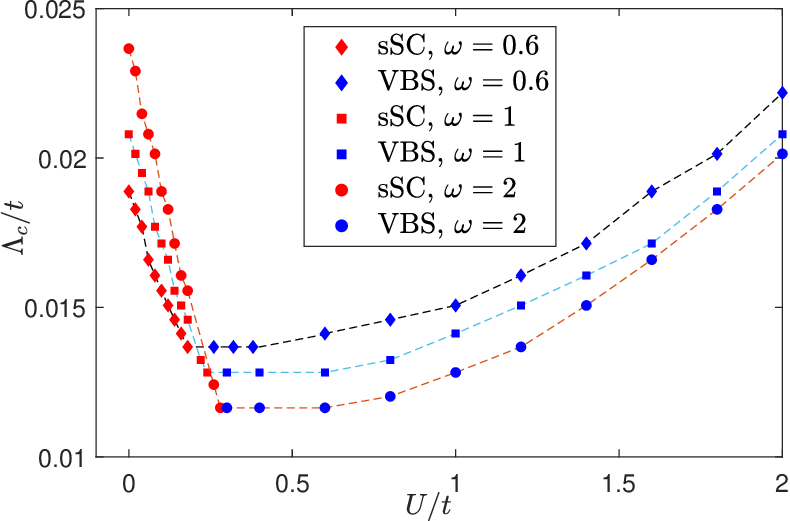}
		\caption{SM-FRG results of the critical energy scales $\Lambda_c$ versus the Hubbard $U$ at $\lambda=0.05$ for several values of $\w$ with the red/blue markers denoting sSC/VBS, respectively.}
		\label{fig:u}
	\end{figure}

We now add the Hubbard interaction $U\sum_in_{i\up}n_{i\dn}$ into the model, to see whether the sSC or VBS survives. Since the DQMC suffers from the negative sign problem in the presence of a finite $U$, we limit ourselves in the SM-FRG calculations.
In Fig.~\ref{fig:u}, the SM-FRG results of the critical energy scale $\Lambda_c$ are plotted versus the Hubbard $U$ at $\lambda=0.05$ for three values of $\w=0.6,1,2$, respectively.
For all these cases, the Hubbard $U$ is found to suppress the sSC quickly and drives the system into the VBS state.
This behavior is because a repulsive $U$ suppresses the onsite pairing.
On the other hand, a large $U$ enhances the charge bond fluctuations and thus favors the VBS state, which is consistent with the previous study \cite{wang2013competing,ChenLH}.
Within the regime in this study, we find no SDW-like instability. The reason is well-known: the matrix element effect almost eliminates the effect of $U$ in the spin scattering envolving the same sublattice. Interestingly the same matrix element effect is absent in the VBS state since it involves fermions on unequal sublattice.

\subsection{Effect of doping} \label{vho}

\begin{figure}
	\includegraphics[width=\linewidth]{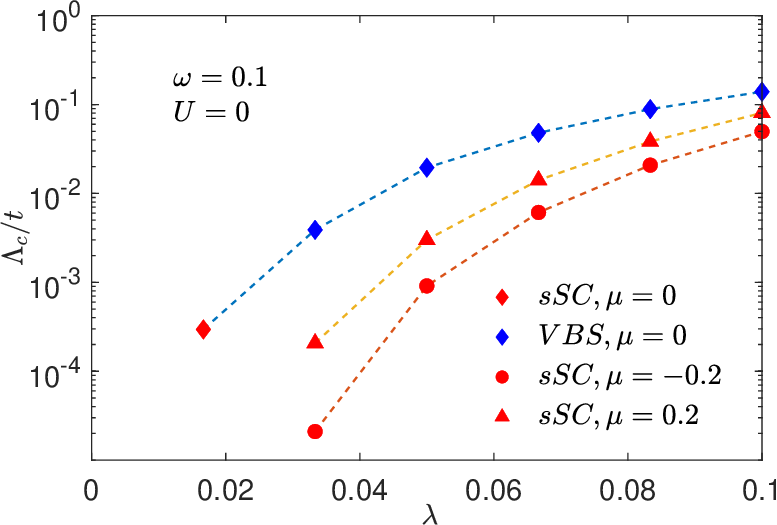}
		\caption{SM-FRG results of the critical energy scales $\Lambda_c$ for sSC (red) and VBS (blue) versus $\lambda$ for $\mu=0$ (diamonds), $0.2$ (triangles) and $-0.2$ (circles), respectively. In the calculations, we have set $\w=0.1$ and $U=0$.}
		\label{fig:vho}
	\end{figure}

Besides the Hubbard interaction U, doping provides another way to affect the competition between the sSC and VBS orders. We consider $\mu=\pm0.2$ slight away from the upper van Hove filling level as shown in Fig.~\ref{fig:band}(b). For these two fillings, the Fermi surfaces are plotted in Fig.~\ref{fig:band}(d), where the electron-like pocket around $K$ corresponds to $\mu=-0.2$ and the hole-like pocket around $\Gamma$ corresponds to $\mu=0.2$.
It is clear that the perfect nesting is no longer present for $\mu\neq 0$.
Therefore, the VBS order is less favored, leaving room for the Cooper instability which does not rely on nesting at all (although nesting could enhance it).
In Fig.~\ref{fig:vho}, we present the SM-FRG results of critical energy scale $\Lambda_c$ versus $\lambda$ at $U=0$, $\w=0.1$ for $\mu=0,\pm0.2$.
At $\mu=0$, an increasing $\lambda$ drives the system from the sSC state to the VBS state.
But for both $\mu=\pm0.2$, the VBS is not present, while
sSC survives, although its $\Lambda_c$ is reduced by the decrease of the density of states.

\section{Summary and Discussions} \label{sec:summary}

In summary, we have performed a systematic study of the ground state properties of the optical SSH model on the kagome lattice at and around the upper van Hove filling. By combining the SM-FRG and projector DQMC methods, at the upper van Hove filling, we find that for a given electron-phonon coupling, the higher-frequency bond phonon induces the sSC state while the lower-frequency phonon favors the VBS state. The transition frequency $\w_c$ is found to be roughly linear in the electron-phonon coupling. After the Hubbard $U$ is turned on, our SM-FRG results show that the sSC is strongly suppressed, and the VBS is the dominating phase. On the other hand, when the system is doped away from the van Hove filing, the absence of the perfect nesting suppresses VBS in favor of  sSC state.

{
Finally, some remarks are given.
(1) For the layered kagome superconductors AV$_3$Sb$_5$ (A=K, Rb, Cs), the SSH phonon can arise from the up-and-down vibration of the out-of-plane Sb atoms, as shown in the first-principle study \cite{DFT_PRB_2023}.
}
(2) The CBO has been discovered near 90K in AV$_3$Sb$_5$ \cite{k-t-1,k-t-2,k-t-3,k-t-4,Jiang_NSR_2022}, although there is still a debate on whether it is a time-reversal-invariant CBO, or a chiral CBO in the form of loop currents.
(3) Superconductivity in AV$_3$Sb$_5$ is found at lower temperatures, and it is interesting to ask whether and how SC {(either uniform or pair density wave)} and CBO are intrinsically related there. Our results show competition between VBS and sSC, which could be relevant in AV$_3$Sb$_5$.
{
(4) For both the kagome (at VHS filling) and square (at half-filling) lattices, the low frequency SSH phonon favors CBO (or VBS), reminiscent of dimerizations in one dimensional conducting polymers, while the high frequency SSH phonon favors sSC in both systems. It is the additional SU(4) symmetry for the square lattice to further cause the degeneracy among sSC, CDW and AF \cite{yqg2022}.
}

\begin{acknowledgments}
This work is supported by National Key R\&D Program of China (Grant No. 2022YFA1403201) and National Natural Science Foundation of China (Grant No. 12374147, No. 12274205, No. 92365203, and No. 11874205). The numerical calculations were performed at the High Performance Computing Center of Nanjing University.
\end{acknowledgments}


\appendix
\section{SMFRG with retarded interactions} \label{sec:appendixA}
Functional renormalization group (FRG) is one of the powerful methods to study correlated electronic systems.  \cite{Berges_PR_2002,Metzner_RMP_2012,Dupuis_PR_2021,Kopietz__2010}. In this appendix, we mainly focus on one of its realizations, called singular-mode FRG (SM-FRG), in particular with phonon induced retarded interactions as employed in this work.

\subsection{Decomposition in Mandelstam channels and FRG flow equations}
In our SM-FRG, we study the FRG flows of the 4-point one-particle irreducible (1PI) vertices $\Gamma_{1234}$ appearing in the effective interaction  $H_{\Gamma}=\frac{1}{2}\sum_{1,2,3,4}\psi^\dagger_1\psi^\dagger_2\Gamma_{1234}\psi_3\psi_4$, where $\psi$ is the fermion field with the subscript $1,2,3,4$ denoting the one-particle degrees of freedom (such as frequency, momentum, orbital, sublattice, spin, etc.). In the SU(2) symmetric case under concern, the spins for 1 and 4 are the same, and similarly for 2 and 3.

\begin{figure}
	\includegraphics[width=\linewidth]{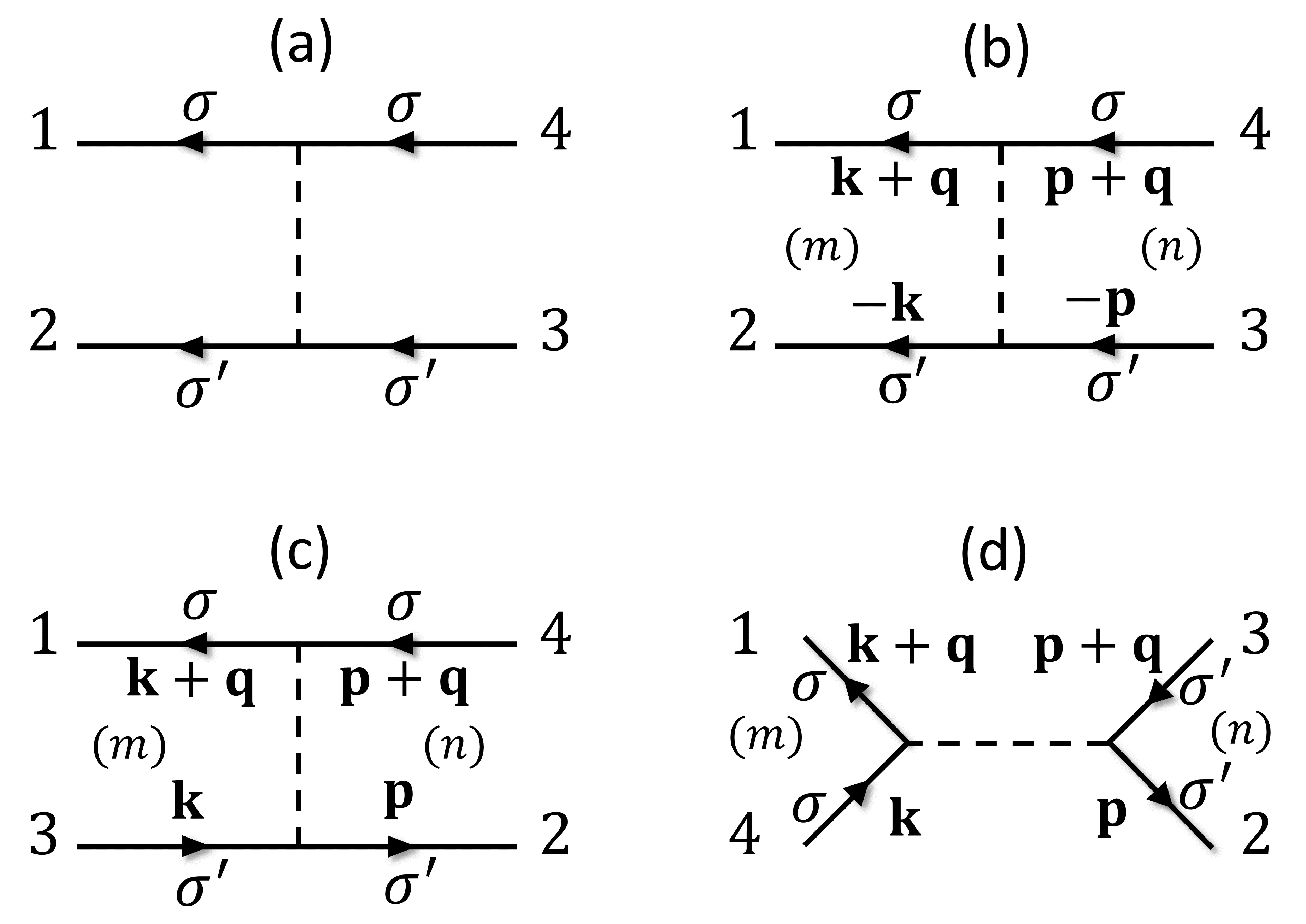}
	\caption{A generic 4-point 1PI vertex (a) can be rearranged into the pairing (P), crossing (C) and direct (D) channels as shown in (b)-(d), respectively.
	The momentum $\mathbf{k,q,p}$ are explicitly shown for clarity. The spins ($\sigma$ and $\sigma'$) are conserved during fermion propagation in the spin-SU(2) symmetric case. The labels $m$ and $n$ denote fermion bilinears.}
	\label{fig:vertex}
\end{figure}

We define fermion bilinears in the three Mandelstam channels as,
\begin{align}
&\alpha_{12}^{\dag}=\psi_1^\dag\psi_2^\dag \ \text{ (pairing)}, \nn\\
&\beta_{13}^{\dag}=\psi_1^\dag\psi_3\  \text{ (crossing)}, \nonumber\\
&\gamma_{14}^{\dag}=\psi_1^\dag\psi_4 \ \text{ (direct)}.
\end{align}
Then the general 1PI vertex can be rewritten as scattering matrices $P$, $C$ and $D$ in the three channels as
\begin{align}
H_\Gamma=&\frac12 \sum_{12,43}\alpha_{12}^{\dag} ~P_{12;43}~\alpha_{43} \nn\\
=& -\frac12 \sum_{13,42}\beta_{13}^{\dag} ~C_{13;42}~ \beta_{42} \nonumber \\
=& \frac12 \sum_{14,32}\gamma_{14}^{\dag} ~D_{14;32} ~\gamma_{32}, \label{eq:rewind}
\end{align}
as illustrated in Fig.~\ref{fig:vertex}(b)-(d).
The collective 4-momentum (of the two fermions in a bilinear) is
$q = k_1+k_2$, $k_1-k_3$, $k_1-k_4$, in the P, C and D channels.
If the subscripts $1,2,3,4$ run over all sites, the three scattering matrices $P$, $C$ and $D$ are all equivalent to $\Gamma$, i.e. $\Gamma_{1234}=P_{12;43} =C_{13;42}=D_{14;32}$.
But in practical calculations, the fermion bilinears must be truncated. On physical grounds, the important bilinears are those that join the singular scattering modes, and such eigen modes determine the emerging order parameter. Since order parameters are composed of short-ranged bilinears, only such bilinears are important. These include onsite and on-bond pairing in the pairing channel, and onsite and on-bond particle-hole density in the C and D channels. The FRG based on the decomposition of the interaction vertices into scattering matrices in the truncated fermion bilinear basis, which are sufficient to capture the most singular scattering modes, is called the singular-mode FRG (SM-FRG). \cite{wang2012functional,xiang2012high,wang2013competing}

\begin{figure}
	\includegraphics[width=0.45\textwidth]{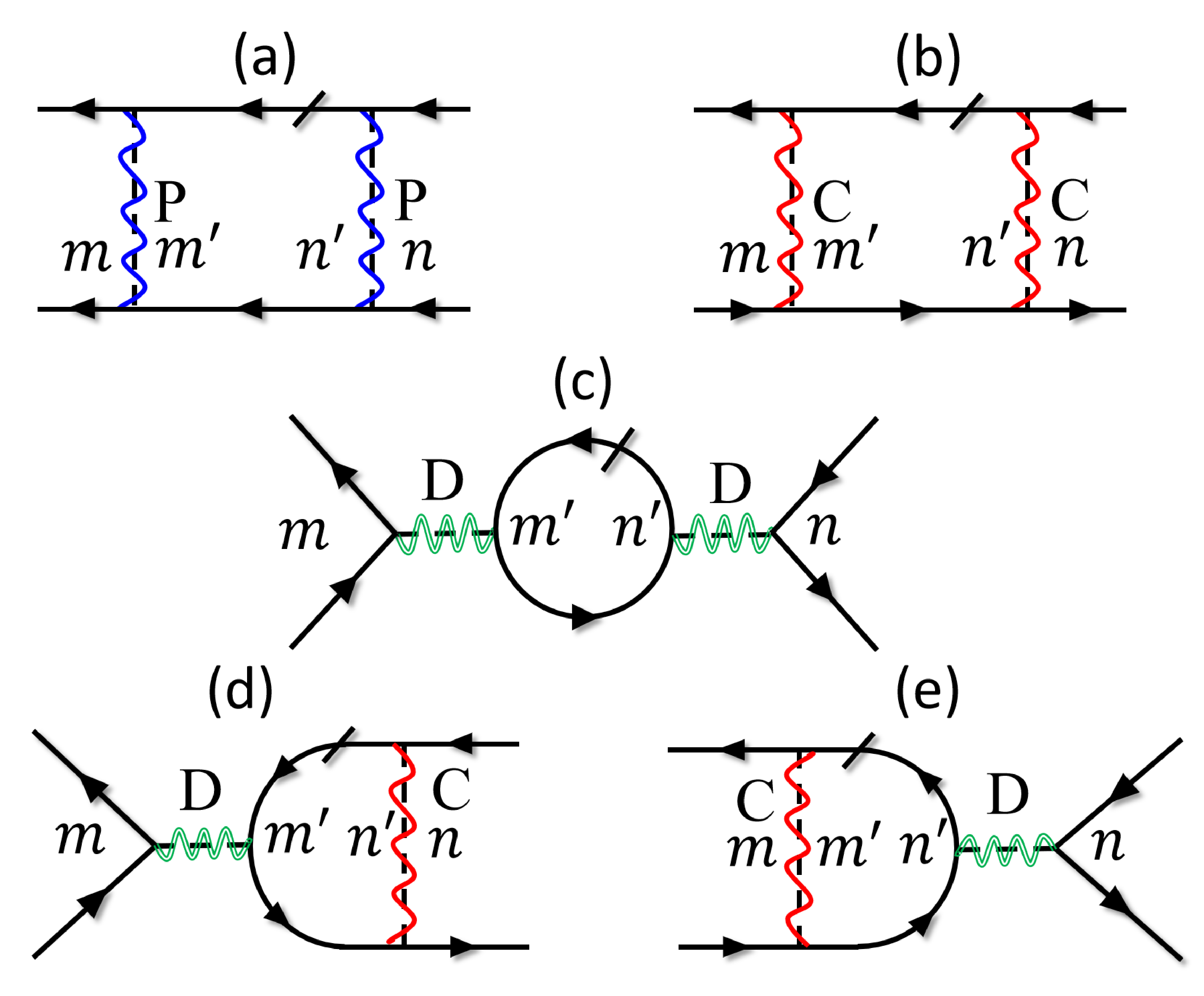}
	\caption{One-loop contributions to $\partial\Gamma_{1234}/\partial\Lambda$. The dashed and wavy line denote the contribution from $\Gamma$ and $\Pi_{\nu}$ respectively. They are added up in the calculation. The slash denotes the single-scale propagator and can be put on either
    one of the fermion lines within the loop.  Note that $\Pi_\nu$ enters at Matsubara frequency $\nu=\Lambda$ (blue and red wavy lines) in P and C channels, while it does at $\nu=0$ (green wavy lines) in the D channel.}
	\label{fig:1loop}
\end{figure}

Starting from $\Lambda=\infty$ where the 1PI vertices $P$, $C$ and $D$ are given by the bare interactions, $\Gamma_{1234}$ flows as
\begin{align} \label{eq:flow}
\frac{\partial\Gamma_{1234}}{\partial\Lambda} =& [P\chi_{pp}P]_{12;43}+[C\chi_{ph}C]_{13;42} \nonumber\\
&+ [D\chi_{ph}C+C\chi_{ph}D-2D\chi_{ph}D]_{14;32} ,
\end{align}
see Fig.~\ref{fig:1loop} for illustration.
The products within the square brackets imply matrix convolutions, and $\chi_{pp}$ and $\chi_{ph}$ are single-scale (at $\Lambda$) susceptibilities given by, in real space,
\begin{align}
	[\chi_{pp}]_{ab;cd} &= \frac{1}{2\pi}\left[ G_{ac}(i\Lambda)G_{bd}(-i\Lambda)+
	(\Lambda \to -\Lambda) \right], \\
	[\chi_{ph}]_{ab;cd} &= \frac{1}{2\pi}\left[ G_{ac}(i\Lambda)G_{db}(i\Lambda)+
	(\Lambda \to -\Lambda) \right],
\end{align}
where $a,b,c,d$ are dummy fermion indices (that enter the fermion bilinear labels), and $G_{ab}(i\Lambda)$ is the normal state Matsubara Green's function. (The expression in the momentum space is slightly more complicated but is otherwise straightforward, and is used in actual calculations.)
As usual \cite{Metzner_RMP_2012}, we neglect self-energy correction which could be absorbed in the band dispersion, and we also neglect the sixth and higher order vertices, which are RG irrelevant at low energy scales. The frequency dependence of the 4-point vertices is also RG irrelevant and ignored. In this spirit, all external legs are set at zero frequency.
The functional flow equation is solved by numerical integration over $\Lambda$.
Note that after $\Gamma_{1234}$ is updated after an integration step, it is rewinded as $P$, $C$ and $D$ according to $\Gamma_{1234}=P_{12;43} =C_{13;42}=D_{14;32}$.
In this way, SM-FRG can treat interactions in all channels on equal footing. In fact, if we ignore the channel overlaps, the flow equation reduces to the ladder equation in the P-channel, and the random-phase-approximation in the C- and D-channels. The SM-FRG combines the three channels coherently.

\subsection{Including phonon-mediated interactions}
The phonon-mediated interaction can be included as a part of $\Gamma_{1234}$.
In principle, Eq.~\ref{eq:flow} can be directly applied to the total interaction including the retarded one, by keeping the full frequency dependence. However, the frequency-dependence in the FRG-generated correction to the four-point vertices can be argued to be RG irrelevant \cite{Kopietz__2010, Metzner_RMP_2012}. In this spirit, we separate the total vertex $\Gamma$ into an instantaneous part $\Gamma^I$ and a retarded one $\Gamma^R$,
\begin{align}
	\Gamma=\Gamma^I+\Gamma^R.
\end{align}
Since we take the FRG-corrected part as instantaneous, the retarded part is always given by $\Gamma^R=\Pi_\nu$ for the associated fermions. The initial value of $\Gamma^I$ at $\Lambda=\infty$ is given by the Hubbard $U$. For brevity, we will keep using the notations $\Gamma$, $P$, $C$, and $D$ (without superscript ``$I$'') for the instantaneous part.
The Feynman diagrams contributing to the flow of $\Gamma$ is illustrated in Fig.~\ref{fig:1loop}, where the dashed lines are the instantaneous vertices, and the wavy lines are from the retarded kernel suitably added to the instantaneous part. Explicitly, the flow equation can be written as
\begin{align} \label{eq:ph-flow}
\frac{\partial\Gamma_{1234}^I}{\partial\Lambda} &= [\+P\chi_{pp}\+P]_{12;43}+[\+C\chi_{ph}\+C]_{13;42} \nn\\
&+[-2\+D\chi_{ph}\+D + \+D\chi_{ph}\+C + \+C\chi_{ph}\+D]_{14;32} ,
\end{align}
where $\+P=P+[\Pi_\Lambda]_P$, $\+C=C+[\Pi_\Lambda]_C$, $\+D=D+[\Pi_0]_D$, with $[\Pi_\nu]_{P,C,D}$ the projection of the phonon-induced interaction projected in the respective channels (or associated to the desired fermion bilinears).
Note that the Matsubara frequency of $\Pi_\nu$ is $\Lambda$ in $\+P$, $\+C$, and $0$ in $\+D$, as a result of frequency conservation when the external fermions are all set at zero frequency in Fig.~\ref{fig:1loop}.

\subsection{Singular scattering modes and order parameters}
The scattering matrices in the SC, SDW and CDW channels can be shown to be related to $P$, $C$ and $D$ as follows,
\begin{align}
    V^{\rm SC}=P,\ \ V^{\rm SDW}=-C,\ \ V^{\rm CDW}=2D-C.
\end{align}
In a given channel, the matrix can be decomposed by singular value decomposition (SVD), in momentum space,
\begin{align}
V_{mn}(\0q)=\sum_{\alpha} \phi_{\alpha m}(\0q) S_{\alpha}(\0q) \phi^*_{\alpha n}(\0q),
\end{align}
where $m,n$ label the fermion bilinear, $S_\alpha$ and $\phi_{\alpha n}$ are eigen value and vector for the $\alpha$-th singular mode.

During the SM-FRG flow, we monitor the leading (most negative) eigenvalue, which we abbreviate as $S$ in each channel.
As the energy scale $\Lambda$ reduces, the first divergence of $S$ indicates a tendency towards an instability with order parameter described by the associated eigen mode $\phi(\0Q)$, where $\0Q$ is the associated collective momentum. In this case, one can drop the non-singular components to write the renormalized interaction as,
\begin{equation}
H_\Gamma \sim \frac{S}{N}O^\dagger O + \cdots,
\end{equation}
where $N$ is the number of unit cells, $O$ is the mode operator that is a combination of the fermion bilinears (see below), and the dots represent symmetry related terms. For example, if the SC channel diverges first, we have
\begin{align}
O_{\rm SC}^\dagger &= \sum_n \phi_n(\0Q) \alpha_n^\dagger(\0Q) \nonumber \\
	&\to \sum_{\0k,n=(a,b,\delta)}\psi_{\0k+\0Q, a}^\dagger \phi_n(\0Q) e^{i\0k\cdot\delta}\psi_{-\0k,b}^\dagger,
\end{align}
where $n$ labels a fermion bilinear, $a$ and $b$ denote the sublattice, and $\Delta_\delta^{ab}=\phi_n(\0Q)$ is the element of the real-space pairing matrix on the bond $\delta$ radiating from $a$.
The spin indices do not have to be specified, as the symmetry of the gap function under inversion automatically determines whether the pair is in the singlet or triplet state.
Because of the eventual Cooper mechanism, the divergence in the pairing channel always occurs at momentum $\0Q=0$.

Similarly, if the SDW channel diverges first, we obtain the mode operator
\begin{align}
O_{\rm SDW}^\dagger  &= \sum_n \phi_n(\0Q)\beta_n^\dagger(\0Q) \nn \\
&\to \sum_{\0k,n=(a,b,\delta)}\psi_{\0k+\0Q,a,\uparrow}^\dagger \phi_n(\0Q) e^{i\0k\cdot\delta}\psi_{\0k,b,\downarrow},
\end{align}
where we assign the spin order in the transverse direction.

Finally, if the CDW channel diverges first, we obtain the mode operator
\begin{align}
O_{\rm CDW}^\dagger &= \sum_n \phi_n(\0Q)\gamma_n^\dagger(\0Q)\nn \\
&\to \sum_{\0k,\sigma,n=(a,b,\delta)}\psi_{\0k+\0Q,a,\sigma}^\dagger \phi_n(\0Q) e^{i\0k\cdot\delta}\psi_{\0k,b,\sigma},
\end{align}
Note that $H_{\rm SDW/CDW}$ can capture both onsite and on-bond density waves, since the fermion bilinears contain both cases of $\delta=0$ and $\delta\neq 0$.

\bibliography{reference}

\end{document}